\documentclass[10pt,twocolumn,showpacs,preprintnumbers,amsmath,amssymb,aps,superscriptaddress,longbibliography,pra]{revtex4-1}
\usepackage{appendix}
\usepackage{bbm}
\usepackage{mathrsfs}
\usepackage{graphicx}
\usepackage{dcolumn}
\usepackage{bm}
\usepackage{braket}
\usepackage{amsmath}
\usepackage{amsfonts}
\usepackage[dvipsnames]{xcolor}
\usepackage[colorlinks=true,linkcolor=Blue,urlcolor=BlueViolet,citecolor=BlueViolet]{hyperref}
\usepackage{natbib}
\usepackage{multirow}

\begin{document}

\title{Quantum non-local nonstabilizerness}
\author{Dongheng Qian}
\affiliation{State Key Laboratory of Surface Physics and Department of Physics, Fudan University, Shanghai 200433, China}
\affiliation{Shanghai Research Center for Quantum Sciences, Shanghai 201315, China}
\author{Jing Wang}
\thanks{Contact author: wjingphys@fudan.edu.cn}
\affiliation{State Key Laboratory of Surface Physics and Department of Physics, Fudan University, Shanghai 200433, China}
\affiliation{Shanghai Research Center for Quantum Sciences, Shanghai 201315, China}
\affiliation{Institute for Nanoelectronic Devices and Quantum Computing, Fudan University, Shanghai 200433, China}
\affiliation{Hefei National Laboratory, Hefei 230088, China}

\begin{abstract}
Quantum entanglement and quantum nonstabilizerness are fundamental resources that characterize distinct aspects of a quantum state: entanglement reflects non-local correlations, while nonstabilizerness quantifies the deviation from stabilizer states. A quantum state becomes a valuable resource for applications like universal quantum computation only when both quantities are present. Here, we propose that quantum non-local nonstabilizerness (NN) serves as an effective measure of this combined resource, incorporating both entanglement and nonstabilizerness. We demonstrate that NN can be precisely computed for two-qubit pure states, establishing a direct relationship with the entanglement spectrum. We then extend the definition of NN to mixed states and explore its presence in many-body quantum systems. We reveal that the two-point NN decays according to a power law in critical states, indicating that it can serve as an effective quantum information probe for phase transitions. Furthermore, we explore measurement-induced NN and uncover an intriguing phenomenon termed ``nonstabilizerness swapping'', wherein post-measurement NN decays more slowly than any pre-measurement correlations. The occurrence of this phenomenon provides a sufficient condition for both entanglement swapping and the presence of a nontrivial sign structure. Our results thus represent a pivotal step toward accurately quantifying the “quantumness” of a state and bring insights from multiple perspectives into the property of NN.

\end{abstract}

\date{\today}

\maketitle

\section{Introduction}
\label{introduction}
Quantum entanglement, which captures the inability of a quantum state to be decomposed into product states of its subsystems, is a fundamental manifestation of quantum mechanics~\cite{horodecki2009}. It serves as a cornerstone for numerous applications, including quantum computation, quantum simulation, and quantum communication~\cite{nielsen2010, wilde2017}. Entanglement is conventionally viewed as a non-local resource that cannot be generated through local operations alone, and is typically quantified by the von Neumann entropy for pure states or the entanglement negativity for mixed states~\cite{bennett1996, plenio2005, vidal2002, zyczkowski1998, horodecki1996}. Since an entangled state of $N$ qubits generally requires storing $O(2^N)$ complex amplitudes for its classical representation, quantum algorithms leveraging such states can achieve exponential speedups over direct classical simulations, potentially even surpassing any known classical algorithm—a concept referred
to as quantum supremacy or, more recently, quantum advantage~\cite{arute2019,boixo2018,harrow2017}. 

Nevertheless, entanglement alone does not fully capture the ``quantumness'' of a state. Certain highly entangled states, known as stabilizer states, can still be efficiently described and simulated~\cite{gottesman1998,gottesman1997,aaronson2004}. A stabilizer state is uniquely specified by a set of stabilizer generators, which requires only $O(N^2)$ memory space. Furthermore, when a stabilizer state evolves under unitary gates from the Clifford group or under Pauli measurements, it remains within the stabilizer formalism, allowing for efficient classical simulation of the entire process. Consequently, nonstabilizerness, also referred to as ``magic'', is equally crucial for a quantum state to serve as a valuable resource in demonstrating quantum advantage~\cite{beverland2020a,bravyi2016a,bravyi2019a,bravyi2016b,bravyi2005,bu2019,campbell2011,gu2024a,gu2024b,gu2024c,howard2017, liu2022, seddon2021,true2022,yoganathan2019,zhou2020,bejan2024, koh2017, bouland2018, zhang2024}. It is worth emphasizing that a key difference between these two resources is that entanglement remains invariant under local unitary transformations, whereas nonstabilizerness is basis-dependent, implying that it can be altered through local operations.
 
Thus, a natural question arises: Is there a suitable measure of the genuinely non-classical resources that integrates both entanglement and nonstabilizerness? Moreover, is this measure computationally tractable? Theoretically, such a measure would deepen our understanding of the connection between these two seemingly distinct properties~\cite{Iannotti2025a, Szombathy2025}, while practically it would help characterize algorithmic complexity and guide the selection of initial states in quantum computations and simulations. With the existence of such a measure, it becomes pertinent to identify resourceful states~\cite{chitambar2019}. Furthermore, a more intriguing question is whether this resource can be manipulated through measurement.

In this paper, we address these questions by investigating quantum non-local nonstabilizerness (NN), which quantifies the minimal amount of nonstabilizerness contained in a bipartite state after optimizing over all possible local unitary transformations on each subsystem~\cite{cao2024a, cao2024b, Cepollaro2025}. We first demonstrate that NN is a promising candidate for characterizing quantum resources and is applicable to any faithful and additive nonstabilizerness measure. Although previous studies have focused on providing bounds and estimates of NN, we show that NN is analytically tractable for a two-qubit pure state and directly related to the entanglement spectrum. We then investigate which states exhibit nonzero NN. Our results indicate that NN is ubiquitous in typical two-qubit pure states, with Haar-random states having the highest probability of attaining maximal NN. Extending the definition to the mixed state, we further examine the behavior of two-point NN in critical many-body states, using both the transverse-field Ising model (TFIM) at the critical point and a monitored Haar-random unitary circuit (MHC) with critical measurement rate as representative examples~\cite{li2019, li2021, zabalo2020}. We find that two-point NN decays following a power law, mirroring the behavior of mutual information. These findings establishes NN as an additional, effective probe for phase transitions from a quantum information perspective.
Finally, we explore whether NN can be manipulated through measurements, akin to the manipulation of entanglement. Intriguingly, measurement-induced NN in a critical MHC decays more slowly than any correlation present before measurement, revealing a ``nonstabilizerness swapping'' phenomenon reminiscent of entanglement swapping~\cite{audenaert2005a, bennett1993}.

\section{Non-local nonstabilizerness}
\label{nonlocal_NN}
\begin{figure}[t]
\begin{center}
\includegraphics[width=3.4in, clip=true]{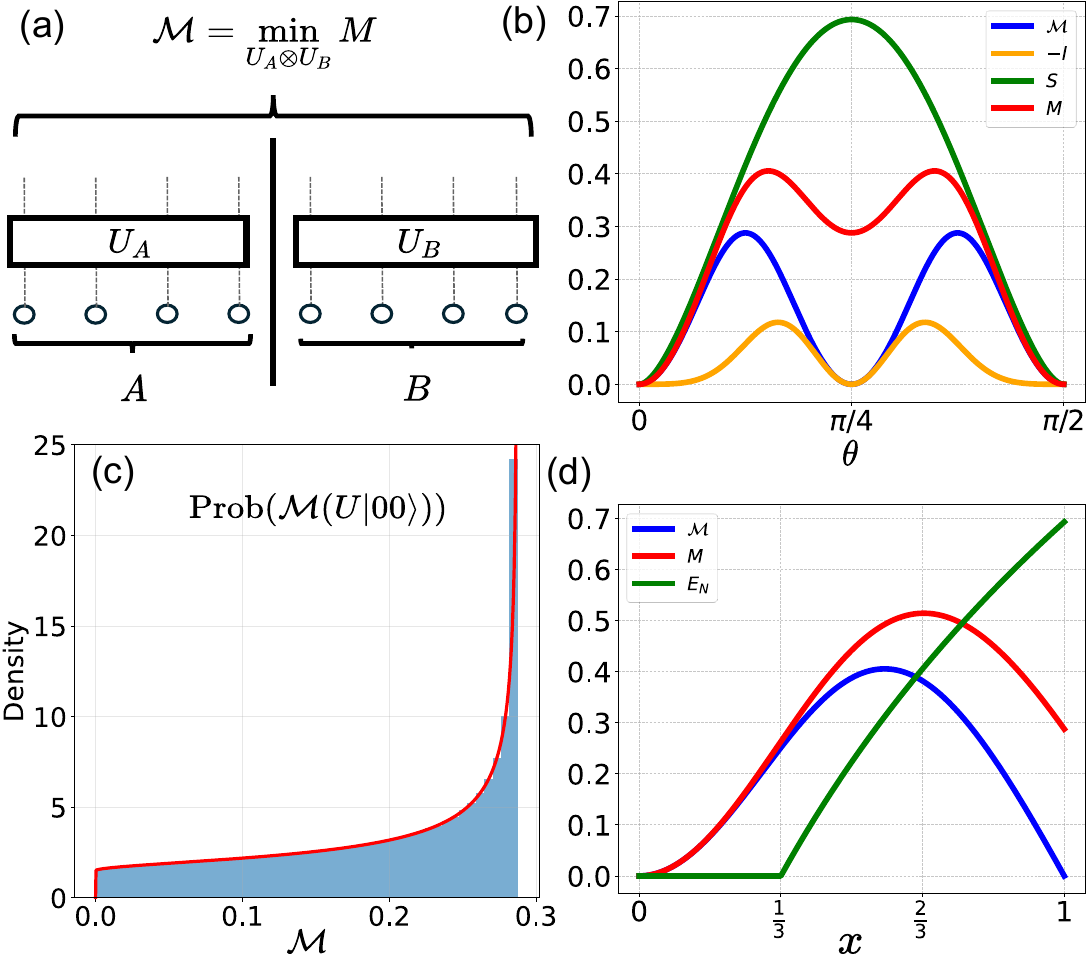}
\end{center}
\caption{Definition of NN and two-qubit NN. (a) An illustration of how NN is defined. (b) A comparison of NN $\mathcal{M}$, von Neumann entropy $S$, and mutual SRE $I$ for the state $\text{cos}(\theta)\left | 00 \right \rangle + \text{sin}(\theta)\left | 11 \right \rangle$. Notice that $I$ is negative. As $\theta$ approaches $0$ and $\pi /2$, the state becomes a product state, while at $\theta=\pi/4$, it is a stabilizer state. The total nonstabilizerness $M$ is also computed after applying a $\text{T}$-gate to the first qubit, where $\mathcal{M}$ remains unchanged. (c) The probability distribution of $\mathcal{M}$ for random pure states, with the red line representing the analytical result and the histogram depicting the numerical computation. (d) A comparison of $\mathcal{M}$ and logarithmic negativity $E_{N}$ for the Werner state. The total nonstabilizerness $M$ after applying a $\text{T}$-gate on the first qubit is also calculated. For the Werner state, the total nonstabilizerness $M$ always equals to the mutual SRE $I$. }
\label{fig1}
\end{figure}

We consider a bipartite pure state $\left | \Psi_{AB} \right \rangle$, where $A$ and $B$ represent two subsystems. As illustrated in Fig.~\ref{fig1}(a), the NN is defined as~\cite{cao2024a, cao2024b}:
\begin{equation}
\label{eq:1}
\mathcal{M}_{AB}(\left | \Psi_{AB} \right \rangle) = \min_{U_{A}\otimes U_B} M(U_{A}\otimes U_B\left | \Psi_{AB} \right \rangle), 
\end{equation}
where $M$ is a measure of nonstabilizerness, and $U_{A}$ and $U_{B}$ are unitary gates acting on subsystems $A$ and $B$, respectively. We will omit the subscript of $\mathcal{M}_{AB}$ when the bipartition is clear from the context. In operational terms, a finite NN indicates that the state cannot be efficiently represented as a stabilizer state, regardless of the local unitaries applied to subsystems $A$ and $B$ in an effort to reduce the nonstabilizerness. By definition, $\mathcal{M}$ is symmetric under the exchange of subsystems $A$ and $B$ and is stable under local basis rotations, which can be regarded as free operations.  
Since $\mathcal{M}$ can be directly related to the flatness of the entanglement spectrum, a nonzero value of $\mathcal{M}$ directly implies that the state possesses both entanglement and nonstabilizerness~\cite{tirrito2024}. For any faithful nonstabilizerness measure $M$, $\mathcal{M}$ would be further guaranteed to be non-negative. 
Moreover, if $M$ is additive, i.e., $M(\left | \psi \right \rangle \otimes \left | \phi \right \rangle) = M(\left | \psi \right \rangle) + M(\left | \phi \right \rangle)$, then the NN is sub-additive. Specifically, for two bipartite pure states $\left | \Psi_{A_1B_1} \right \rangle$ and $\left | \Phi_{A_2B_2} \right \rangle$, we have $\mathcal{M}_{A_1B_1}(\left | \Psi_{A_1B_1} \right \rangle) + \mathcal{M}_{A_2B_2}(\left | \Phi_{A_2B_2} \right \rangle) \geq \mathcal{M}_{AB}(\left | \Psi_{A_1B_1} \right \rangle \otimes \left | \Phi_{A_2B_2} \right \rangle)$, where $A = A_1 \cup A_2$ and $B = B_1 \cup B_2$. The proof can be found in Appendix~\ref{sub-additivity}. In summary, for any additive and faithful nonstabilizerness measure, the NN is stable, sub-additive, symmetric and vanishes for both product states and stabilizer states. Given these desirable properties and clear physical significance, $\mathcal{M}$ serves as an effective measure to characterize the novel resources of a quantum state that account for both nonstabilizerness and entanglement.

Although various nonstabilizerness measures satisfy faithfulness and additivity, we select $M$ to be the second stabilizer R\'enyi entropy (SRE) in the following, as it is well defined for qubits and avoids any minimization procedure in its calculation~\cite{leone2022, haug2024a, lami2023, leone2024a, leone2024b, niroula2024, odavic2023, oliviero2022a, oliviero2022b,rattacaso2023,tarabunga2024a, tarabunga2023,frau2024, lopez2024a, li2024, fux2024, ding2025}. It is given by:
\begin{equation}
M(\rho) = - \text{ln}\left(\frac{\sum_{P\in\mathbb{P}_N}|\text{Tr}(\rho P)|^4}{\sum_{P\in\mathbb{P}_N}|\text{Tr}(\rho P)|^2}\right), 
\end{equation}
which for pure states reduces to $M(\left | \psi \right \rangle) = - \text{ln}\sum_{P\in\mathbb{P}_{N}}(1/2^{N})|\left \langle \psi  |P|\psi \right \rangle|^4 $, where $N$ is the number of qubits and $\mathbb{P}_N$ represents the set of $N$-fold tensor products of Pauli matrices, denoted as $\sigma_{\mu}$ with $\mu = 0, x, y, z$. It is also instructive to compare $\mathcal{M}$ with the mutual SRE introduced in previous work, which is defined as $I_{AB}(\left | \Psi_{AB} \right \rangle) = M(\left | \Psi_{AB} \right \rangle) -  M(\rho_{A})-M(\rho_B)$, where $\rho_A$ and $\rho_B$ are reduced states of subsystems $A$ and $B$, respectively~\cite{tarabunga2023, lopez2024a, frau2024}. While the mutual SRE captures certain aspects of the non-locality of a state's nonstabilizerness, it lacks stability under local unitary transformations~\cite{fliss2021, haug2023a}, can take negative values in some cases, and generally does not align with the NN $\mathcal{M}$, as we will demonstrate later.

\section{Two-qubit NN}
\label{2qubit_NN}
In general, computing $\mathcal{M}$ is highly complicated because it involves minimizing over the entire spectrum of local unitary transformations.
However, for the two-qubit case, we demonstrate that an analytic expression for $\mathcal{M}$ can be derived. We first prove the following theorem.

\textbf{Theorem:} For a two-qubit pure state of the form $\left | \psi \right \rangle = \text{cos}(\theta)\left | 00 \right \rangle + \text{sin}(\theta)\left | 11 \right \rangle $, $\mathcal{M}(\left | \psi \right \rangle) = M(\left | \psi \right \rangle) = \text{ln}(\frac{8}{7+\text{cos}(8\theta)})$. 

The proof is provided in Appendix~\ref{proof_theorem}, where the key idea is that the Pauli basis representation of this specific state is in a canonical form, with the reduced density matrices of both qubits exhibiting fully polarized Bloch vectors, and the full state having a diagonal correlation matrix~\cite{verstraete2002, jevtic2014}. The following important corollary arises: 

\textbf{Corollary: } For any two-qubit pure state $\left | \Psi \right \rangle$ with the bipartite entanglement spectrum $\{\text{cos}^2(\theta), \text{sin}^2(\theta)\}$, $\mathcal{M}(\left | \Psi \right \rangle) = \text{ln}(\frac{8}{7+\text{cos}(8\theta)})$. 

This holds because any two-qubit pure state can be transformed into the form $\left | \psi \right \rangle = \text{cos}(\theta)\left | 00 \right \rangle + \text{sin}(\theta)\left | 11 \right \rangle $ by applying an appropriate local basis rotation, which can be determined via singular value decomposition. As a result, the NN is directly related to the entanglement spectrum and can be efficiently computed.
It should be noted that NN has previously been understood to be approximated by quantities derived from the entanglement spectrum~\cite{cao2024a}. Our results show that this approximation is, in fact, exact for the two-qubit case.
A direct implication is that any two-qubit pure state can have at most $\text{log}(4/3)$ amount of NN, which equals the nonstabilizerness of the H-type magic state~\cite{bravyi2005}. 
Additionally, it is noteworthy that, when combined with the sub-additivity of $\mathcal{M}$, if a many-body state can be decomposed into product states of multiple two-qubit states of the above form, an upper bound for its NN can be easily determined.

We compute $\mathcal{M}(\text{cos}(\theta)\left | 00 \right \rangle + \text{sin}(\theta)\left | 11 \right \rangle)$ for different values of $\theta$ and compare it with the von Neumann entropy $S$ and the mutual SRE $I$. As shown in Fig.~\ref{fig1}(b), $\mathcal{M}$ vanishes for both stabilizer states and product states, while $S$ solely quantifies the entanglement of the state. Although the mutual SRE $I$ exhibits a similar trend to $\mathcal{M}$, it takes negative values and its peak differs from that of $\mathcal{M}$. Notably, while for this specific form of state, $\mathcal{M}$ is equal to the total nonstabilizerness $M$, differences arise when a local gate is applied to either qubit. For instance, when a $\text{T}$-gate is applied to the first qubit, the NN remains unchanged, but the total nonstabilizerness $M$ changes significantly, displaying a nonzero value at $\theta=\pi/4$. 

Subsequently, we investigate the typical amount of NN in a random two-qubit pure state. We generate random pure states via a Haar-random two-qubit unitary gate and examine the distribution of $\mathcal{M}$ across these states. The results, shown in Fig.~\ref{fig1}(c), indicate that the state most commonly hosts the maximal $\mathcal{M}$. While the analytical expression for this probability distribution can be derived in Appendix~\ref{probability_distribution}, a closed-form expression for the average value of $\mathcal{M}$, analogous to that for entanglement~\cite{page1993,sen1996}, remains unclear. Nevertheless, numerical results suggest that the average $\mathcal{M}$ is approximately $0.192$.

\section{Mixed state generalization}
\label{mixed_state}
To explore NN in a many-body state, it is first essential to extend the definition of NN to mixed states, with particular attention to the two-qubit case.
For mixed states, the NN can be defined analogously to Eq.~(\ref{eq:1}) as $\mathcal{M}_{AB}(\rho_{AB}) = \min_{U_A \otimes U_B}M((U_A \otimes U_B)\rho_{AB}(U_A^{} \otimes U_B)^{\dagger})$. Under this definition, the NN retains the properties of stability, symmetry and subadditivity for any additive measure of nonstabilizerness. By continuing to employ the second SRE, which is faithful when the mixed stabilizer states are regarded resource-free~\cite{mixed}, the NN also vanishes for the mixed stabilizer states~\cite{leone2022, audenaert2005a}. However, a crucial difference is that the NN does not vanish for certain separable states. This difference arises because, even for a single-qubit mixed state, such as $\rho = \frac{1}{2}(\sigma_0 + \frac{1}{2} \sigma_x)$, no unitary transformation can bring it back to a mixed stabilizer state. We also explore other possible nonstabilizerness measures for the mixed state, while the conclusion remains the same (see Appendix~\ref{separable_mixed}).   

Focusing on the two-qubit case, it follows directly from the proof of the previous theorem that if a two-qubit mixed state $\rho$ has a Pauli basis representation in canonical form, we would have $\mathcal{M}(\rho) = M(\rho)$. However, it is important to note that not all mixed states can be transformed into the canonical form using only local unitary transformations~\cite{verstraete2002}. In such cases, numerical optimization is inevitable, as examples will be shown in the next section. Consider the Werner state as an analytically tractable example, which is defined as $\rho_W = x \left | \psi_{-} \right \rangle \left \langle \psi_{-} \right | + (1-x)\mathbb{I}/4$, with $|\psi_{-}\rangle\equiv(|01\rangle-|10\rangle)/\sqrt{2}$~\cite{werner1989}. This state has vanishing Bloch vectors for each qubit and a diagonal correlation matrix $T$ with diagonal entries given by $-x/4$. We calculate NN and logarithmic negativity $E_N$, a measure of separability, with the results shown in Fig.~\ref{fig1}(d). Although the Werner state remains separable for $x < 1/3$, the NN takes on a finite value for small $x$. As $x$ approaches $1$, the state becomes more entangled, yet the NN approaches zero as the state becomes a stabilizer state. The difference between $\mathcal{M}$ and total nonstabilizerness $M$ again becomes evident if we further apply a $\text{T}$-gate on the first qubit. While the NN remain unaffected, the total nonstabilizerness $M$ would change significantly. 

\begin{figure}[t]
\begin{center}
\includegraphics[width=3.4in, clip=true]{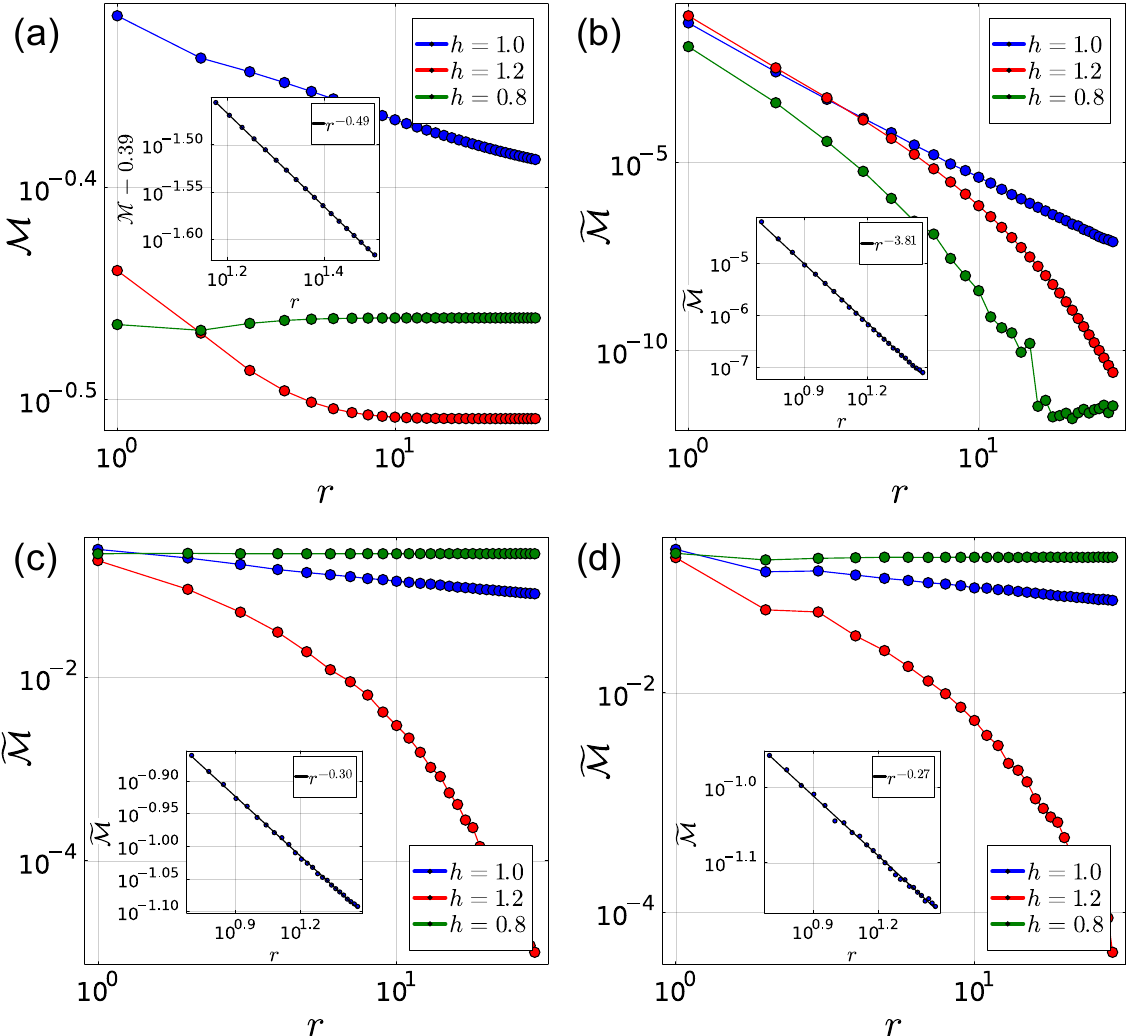}
\end{center}
\caption{Results for the TFIM with system size $L=128$. The ground states are obtained using density matrix renormalization group method, with a maximum bond dimension $\chi=500$. (a) Two-point NN for different $h$. The inset displays a power-law fit with a constant shift for $h=1.0$. (b-d) MINN for measurements along the $x-$, $y-$ and $z-$axes, respectively. The insets show the power-law fit for $h=1.0$. }
\label{fig2}
\end{figure}

\section{NN in many-body states}
\label{NN_many_body}
Building on our improved understanding of the NN in the two-qubit case, we now shift our focus to the scaling behavior of the two-point NN in many-body quantum states, defined as $\mathcal{M}(\rho_{1, r})$, where $\rho_{1,r}$ represents the reduced density matrix for the first qubit and the $r$-th qubit. For systems with translational symmetry and periodic boundary conditions, the two-point NN depends solely on the relative separation $r$. The scaling behavior provides valuable information on the properties of many-body states, especially for critical states. 
We also explore the measurement-induced non-local nonstabilizerness (MINN), defined as $\widetilde{\mathcal{M}}(r) \equiv \sum_i p_i \mathcal{M}(\left | \Psi_{1, r}^{i} \right \rangle)$, where $p_i$ is the probability of obtaining the outcome $i$ from projective measurements performed on all qubits except the first and the $r$-th qubits, and $\left | \Psi_{1, r}^{i} \right \rangle $ represents the corresponding post-measurement state. 
Numerical details are provided in Appendix~\ref{numerical}.

We first consider the TFIM, with Hamiltonian given by $H_{\text{TFIM}} = - \sum_i \sigma_x^{i} \sigma_x^{i+1} - h \sum_i \sigma_z^{i} $. The ground state is paramagnetic for $h > 1$ and ferromagnetic for $h < 1$, with a quantum phase transition occurring at $h=1$. Due to the $\mathbb{Z}_2$ symmetry $\prod_{i}\sigma_{x}^{i}$, the Pauli basis representation of $\rho_{1, r}$ is constrained to be in the canonical form, enabling direct computation of the NN. As shown in Fig.~\ref{fig2}(a), the NN exhibits a power-law decay with an exponent of around $0.5$, close to that of mutual information. In Appendix~\ref{decay_exponent}, we show that their decay exponents are analytically identical for sufficiently large $r$. Away from criticality, the NN decays exponentially to a constant value, distinguishing it from critical states. 
Next, we examine the MINN for different measurement directions. As illustrated in Fig.~\ref{fig2}(b-d), the MINN exhibits a power-law decay for critical states, with different decay exponents depending on the measurement axis. The MINN under $y-$ and $z-$axes measurements decays more slowly than the pre-measurement behavior, but still faster than the slowest decaying correlation prior to measurement, namely the $\left \langle \sigma_x^{1} \sigma_x^{r} \right \rangle$, which decays with exponent $1/4$~\cite{francesco2012}. For non-critical states, the MINN either becomes constant or decays exponentially. These findings indicate that both NN and MINN can serve as effective probes of the underlying quantum phase transitions in many-body systems, in line with the growing interest in probing quantum phase transitions through quantum information measures. 

\begin{figure}[t]
\begin{center}
\includegraphics[width=3.4in, clip=true]{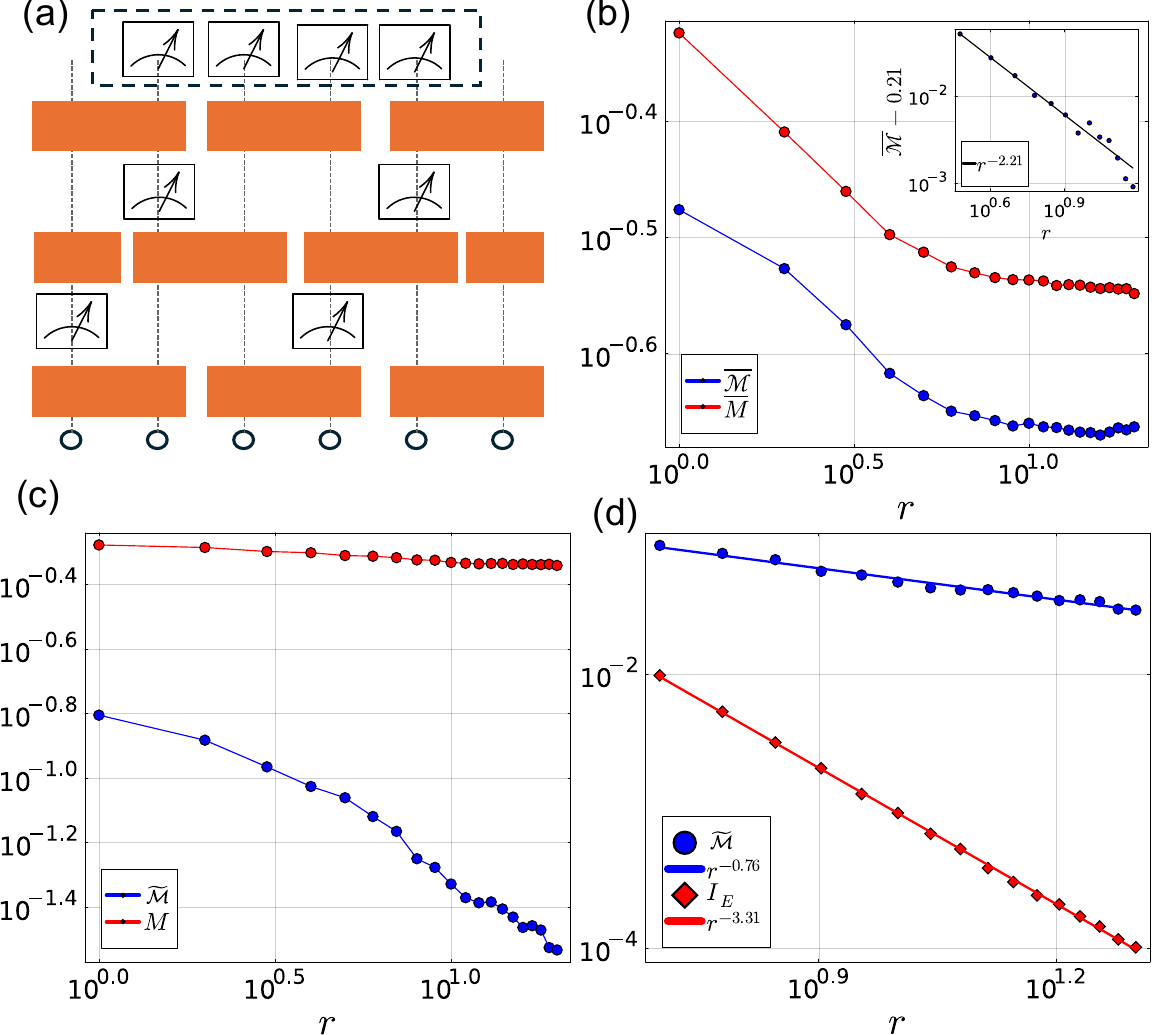}
\end{center}
\caption{Results for critical MHC with $L=48$, with a maximum bond dimension $\chi=1000$. (a) Haar-random unitaries are applied in a brick-wall manner, with a measurement rate of $p=0.17$~\cite{zabalo2020}. The measurements within the dotted box are used solely for computing MINN, with all qubits except the first and the $r$-th qubits are projectively measured along the $z$-axis. (b) A comparison between averaged two-point NN $\overline{\mathcal{M}}$ and total nonstabilizerness $\overline{M}$. Each data point is averaged over 3000 trajectories. The inset shows the power-law fit with a constant shift for $\overline{\mathcal{M}}$. (c) The MINN and the average total nonstabilizerness after measurement, with each data point averaged over $4\times10^4$ trajectories. (d) A comparison between the scaling behavior of the MINN and the mutual information $I_E$ prior measurement. The decay exponents for $\widetilde{\mathcal{M}}$ and $I_E$ are around $0.76$ and $3.31$, respectively. }
\label{fig3}
\end{figure}

We then consider the NN and MINN in MHC at a critical measurement rate of $p=0.17$. The circuit structure is shown in Fig.~\ref{fig3}(a). We consider the average NN defined as $ \overline{\mathcal{M}} = \mathbb{E}_{U} \mathbb{E}_m[\mathcal{M}]$, where $\mathbb{E}_U$ and $\mathbb{E}_m$ represent the averaging over unitary gates and the measurement outcomes, respectively. Since the model is structureless, with only locality as a constraint, $\rho_{1, r}$ does not necessarily have a canonical Pauli basis representation, and the NN must be computed by numerical optimization. As shown in Fig.~\ref{fig3}(b), NN is significantly lower than total nonstabilizerness and also exhibits a power-law decay. We then consider the MINN and compare it with the average total nonstabilizerness for post-measurement states, as presented in Fig.~\ref{fig3}(c). The measurement direction is insignificant, and we assume it to be along the $z$-axis. It is evident that a greater amount of nonstabilizerness can be removed by local unitary transformations as the distance increases. Furthermore, we observe that the decay exponent of MINN is less than half of that of mutual information prior to measurement, as shown in Fig.~\ref{fig3}(d). Since the slowest correlation a state can maintain must decay at an exponent at least half that of mutual information~\cite{wolf2008}, this suggests that the measurement induces NN with a longer range than any pre-existing correlations. This phenomenon is akin to entanglement swapping, where measurements induce long-range entanglement between previously uncorrelated qubits. We refer to this effect as ``nonstabilizerness swapping''. 
Moreover, since the entanglement is lower bounded by $\mathcal{M}$~\cite{cao2024a}, the presence of nonstabilizerness swapping serves as a sufficient condition for the occurrence of entanglement swapping. It is also worth noting that the occurrence of entanglement swapping serves as a sufficient condition for a state to possess a nontrivial sign structure; that is, the state must exhibit certain negative amplitudes in the measurement basis~\cite{lin2023}. Consequently, the phenomenon of nonstabilizerness swapping necessarily implies that the underlying state exhibits a nontrivial sign structure—an alternative diagnostic of a state's “quantumness”~\cite{Hastings2016}.

\section{Discussions}
\label{discussion}
Our results carry several implications across different fields. First, an exact expression for computing the two-qubit NN is not only valuable for quantifying non-classical resources and classical simulation complexity, but also holds direct relevance to the gravitational back-reaction~\cite{cao2024a, cao2024b}. 
Second, the distinct power-law behavior exhibited by the two-point NN at the critical point provides insight into the scaling of NN in generic conformal field theories. Importantly, NN is invariant under local unitary transformations, indicating that it may be computed within the conformal field theory framework in the continuum limit—analogous to the calculation of entanglement entropy. Our exact results for the two-point case thus serve as a benchmark for such a formalism and we leave it to future work. In contrast, mutual SRE is sensitive to microscopic details, making its continuum description more elusive. Third, nonstabilizerness swapping provides a unified manifestation of the three fundamental quantum resource aspects—entanglement, nonstabilizerness, and sign structure—such that any state exhibiting this phenomenon is necessarily nontrivial from all three perspectives. Identifying further examples of such states remains an intriguing open problem, which we leave for future work.

It is noteworthy that a related measure, nonstabilizerness entanglement entropy (NsEE), has recently been introduced~\cite{qian2024, huang2024}. In this formulation, free operations are defined as global Clifford gates instead of local unitary gates, and the computed quantity is the entanglement entropy of the total state rather than its nonstabilizerness. While NsEE and NN aim to characterize the ``quantumness'' of a system from distinct perspectives, further investigation into the relationship between these two measures would be a valuable direction for future research.

\begin{acknowledgments}
We thank Xiaoyu Dong and Pengfei Zhang for valuable discussions. This work is supported by the Natural Science Foundation of China through Grants No.~12350404 and No.~12174066, the Innovation Program for Quantum Science and Technology through Grant No.~2021ZD0302600, the Science and Technology Commission of Shanghai Municipality under Grants No.~23JC1400600, No.~24LZ1400100 and No.~2019SHZDZX01, and is sponsored by ``Shuguang Program'' supported by Shanghai Education Development Foundation and Shanghai Municipal Education Commission.
\end{acknowledgments}

\begin{appendix}

\section{Proof of sub-additivity}
\label{sub-additivity}
In this section, we prove that the NN, defined as $ \mathcal{M}_{AB}(\left | \Psi_{AB} \right \rangle) = \min_{U_{A}\otimes U_B} M(U_{A}\otimes U_B\left | \Psi_{AB} \right \rangle)$, is sub-additive if the nonstabilizerness measure $M$ is additive. By definition, we have:
\begin{equation}
\begin{aligned}
&\mathcal{M}_{AB}(\left | \Psi_{A_1B_1} \right \rangle \otimes \left | \Phi_{A_2B_2} \right \rangle) \\
&= \min_{{U_{A}\otimes U_B}}M(\left | \Psi_{A_1B_1} \right \rangle 
\otimes 
\left | \Phi_{A_2B_2} \right \rangle) \\ 
&= \min_{U_{A}\otimes U_B}[M(\left | \Psi_{A_1B_1} \right \rangle) + M(\left | \Phi_{A_2B_2} \right \rangle)].
\end{aligned}
\end{equation}
The second equation follows directly from the additivity of $M$ and we denote $A = A_1 + A_2$ and $B = B_1 + B_2$. Since unitaries of the form $U_{A_1}\otimes U_{A_2} \otimes U_{B_1} \otimes U_{B_2}$ form a subset of the set of unitaries of the form $U_{A} \otimes U_{B}$, we further have:
\begin{equation}
\begin{aligned}
&\mathcal{M}_{AB}(\left | \Psi_{A_1B_1} \right \rangle \otimes \left | \Phi_{A_2B_2} \right \rangle)  \\
&\leq \min_{U_{A_1}\otimes U_{B_1}}M(\left | \Psi_{A_1B_1} \right \rangle) + \min_{U_{A_2}\otimes U_{B_2}}M(\left | \Phi_{A_2B_2} \right \rangle) \\
&= \mathcal{M}_{A_1B_1}(\left | \Psi_{A_1B_1} \right \rangle) + \mathcal{M}_{A_2B_2}(\left | \Phi_{A_2B_2} \right \rangle).
\end{aligned}
\end{equation}
This establishes the sub-additivity of $\mathcal{M}$.

Building on this sub-additivity, it is straightforward to establish an upper bound for the NN of a many-body state, provided that it can be brought into product states of two-qubit pure states in canonical form via local unitary transformations. It has been shown that such transformations can be approximately realized using the Petz recovery map for Haar-random states~\cite{yoshida2017}. However, the exact determination of such a transformation remains an open problem and is a subject worth further exploration.

\section{Proof of theorem}
\label{proof_theorem}
In this section, we provide a detailed proof of the theorem introduced in the main text.  Any two-qubit state $\rho$ consisting of qubits $A$ and $B$ can be expressed in Pauli basis as:
\begin{equation}
\rho = \frac{1}{4} \sum_{\mu, \nu}R_{\mu\nu}\sigma_{\mu}\sigma_{\nu},
\end{equation}
where $\mu, \nu \in \{0, x, y, z\}$, and the coefficients $R$ are determined by $R_{\mu \nu} = \text{Tr}(\rho \sigma_{\mu} \otimes \sigma_{\nu})$~\cite{schlienz1995, horodecki1996}. The matrix $R$ takes a block structure of the form:
\begin{equation}
  R = \begin{pmatrix}
  1 & \textbf{a}^{T} \\ \textbf{b} & T
  \end{pmatrix},
\end{equation}
where $\textbf{a}$ and $\textbf{b}$ are the Bloch vectors corresponding to the reduced density matrices of qubit $A$ and $B$, respectively, and $T$ represents the correlation matrix. In this representation, the second SRE of the state $\rho$ is concisely expressed as:
\begin{equation}
M(\rho) = -\text{ln}\left(\sum_{\mu\nu}{|R_{\mu \nu}|^4}\right) - S_2(\rho),
\end{equation}
where $S_2(\rho)$ denotes the 2-R\'enyi entropy of the state $\rho$, which is invariant under any unitary transformation applied to $\rho$. Therefore, the calculation of NN $\mathcal{M}$ reduces to finding the appropriate local unitary transformation that maximizes the sum $\sum_{\mu\nu}{|R_{\mu\nu}|^4}$.

Any local unitary transformation of the form $U = U_A \otimes U_B$ corresponds to a basis rotation, resulting in the transformations $\textbf{a} \rightarrow O_A \textbf{a}$, $\textbf{b} \rightarrow O_B \textbf{b}$ and $T \rightarrow O_ATO_B^{T} $, where $O_A$ and $O_B$ are orthogonal matrices corresponding to rotations in qubits $A$ and $B$, respectively. Since $O_A$ is orthogonal, it preserves the norm of the Bloch vector $\textbf{a}$, and we have
\begin{equation}
\sum_i \left | a_i \right |^4 \leq  \sum_i \left | a_i \right |^2 = \text{const.},
\end{equation}
with equality holding if and only if $\textbf{a}$ is either fully polarized or the zero vector. The same statement applies to $\textbf{b}$. Furthermore, if the matrix $T$ is diagonal, i.e., $T = \text{diag}\{d_1, d_2, d_3\}$, the effect of $O_A$ on the $j$-th column of $T$ is governed by  the following inequality: 
\begin{equation}
\begin{aligned}
\sum_{i=1}^{3}\left | T'_{i,j}\right |^4 &= \sum_{i=1}^{3}\left |d_j O_{A,ij}\right | ^4 = \left | d_j \right |^4 \sum_{i=1}^{3}\left | O_{A, ij} \right |^4 \\
&\leq \left | d_j \right |^4  \sum_{i=1}^{3}\left | O_{A,ij} \right |^2 = \left | d_j \right |^4, 
\end{aligned}
\end{equation}
where the final equality follows from the orthogonality of $O_A$. A similar argument holds for the action of $O_B$. This indicates that a diagonal matrix $T$ maximizes the fourth power sum of its elements under any local unitary transformation. Combining these results, we conclude that any local unitary transformation applied to a two-qubit state with fully polarized Bloch vectors for both qubits and a diagonal correlation matrix increases its nonstabilizerness. We call such a state to be in the canonical form of its Pauli basis representation. It follows directly that if a state $\rho$ is in canonical form, then $\mathcal{M}(\rho) = M(\rho)$. 

For the specific state $\left | \psi \right \rangle = \text{cos}(\theta)\left | 00 \right \rangle + \text{sin}(\theta)\left | 11 \right \rangle $, the corresponding $R$ matrix is:
\begin{equation}
R = \begin{pmatrix}
  1& 0 &0  &\text{cos}(2\theta) \\
  0&\text{sin}(2\theta)  &0  &0 \\
  0&  0&  -\text{sin}(2\theta)&0 \\
  \text{cos}(2\theta)& 0 & 0 & 1
\end{pmatrix},
\end{equation}
which clearly exhibits a canonical form. This completes the proof of the theorem stated in the main text.

It is worth noting that while a local unitary transformation always exists that brings a two-qubit pure state into canonical form, this is not generally true for a mixed state. However, a unique normal form exists for any two-qubit mixed state, where the correlation matrix $T$ is diagonal, although the Bloch vectors of the qubits need not be fully polarized~\cite{verstraete2002}.

\section{Probability distribution}
\label{probability_distribution}
In this section, we derive the analytic expression for the probability distribution of NN for the two-qubit Haar random state. It is well known that the Haar random state has an entanglement spectrum $\{\text{cos}(\theta), \text{sin}(\theta)\}$, with the associated probability distribution given by $P(x) = 3(1-2x)^2$, where we define $x \equiv \text{cos}^2(\theta)$~\cite{page1993, sen1996}. From the corollary stated in the main text, the NN for a state with such an entanglement spectrum is expressed as:
\begin{equation}
\begin{aligned}
\mathcal{M}(x) &= \text{ln}\left(\frac{8}{7+\text{cos}(8\theta)}\right) \\
&= \text{ln}\left(\frac{1}{1+4x(x-1)(1-2x)^2}\right).
\end{aligned}
\end{equation}
Our goal is to determine the probability distribution of $y \equiv \mathcal{M}(x)$, where $x$ is a random variable following the distribution $P(x)$. The probability distribution of $y$ is then given by:
\begin{equation}
\label{supp:eq1}
P(y) = \sum_{x:\mathcal{M}(x)=y}\frac{P(x)}{\mathcal{M}'(x)},
\end{equation}
where $\mathcal{M}’(x)$ is the derivative of $\mathcal{M}(x)$ with respect to $x$, which can be explicitly calculated as:
\begin{equation}
\label{supp:eq2}
\mathcal{M}'(x) = -\frac{4(2x-1)(1+8x(x-1))}{1+4(2x-1)^2(x-1)x}.
\end{equation}
Furthermore, the solution to the equation $y = \mathcal{M}(x)$ is:
\begin{equation}
\label{supp:eq3}
x = \frac{1}{2} \pm \frac{1}{4}\sqrt{2 \pm 2e^{-y/2}z}.
\end{equation}
For simplicity, we introduce the notation $z \equiv \sqrt{4-3e^y}$. Substituting Eq.~(\ref{supp:eq2}) and Eq.~(\ref{supp:eq3}) into Eq.~(\ref{supp:eq1}), we finally obtain the probability distribution of $y$ as:
\begin{equation}
\begin{aligned}
P(y) =& \frac{3e^{-y/2}}{4z\sqrt{2(e^y-1)}}\left [(e^{y/2}+z)\sqrt{1-e^{-y/2}z} \right. \\
&+\left . (e^{y/2}-z)\sqrt{1+e^{-y/2}z} \right].
\end{aligned}
\end{equation}
This distribution is plotted in Fig.~\ref{fig1}(c), where it is shown to be consistent with the numerical results. However, we have not been able to find a simple closed-form expression for the average amount of NN, $ \overline{y} \equiv \int yP(y)dy$, for the Haar random state. Consequently, we numerically integrate the expression, which yields an approximate value of $\overline{y}\approx0.1917$.

\section{Decay exponent for TFIM}
\label{decay_exponent}
In this section, we demonstrate that the decay exponent of NN in the critical ground state of TFIM is identical to that of mutual information. We begin by deriving a general result for the decay exponent of the two-point total nonstabilizerness. At the critical point, all connected correlation functions of the form $\mathcal{C}_{\mu\nu} = \text{Tr}(\rho_{1, r} \sigma_{\mu}^{1} \sigma_{\nu}^{r}) - \text{Tr}(\rho_1 \sigma_{\mu}^{1}) \text{Tr}(\rho_r \sigma_{\nu}^{r})$ obey a power-law decay, i.e., $\mathcal{C}_{\mu \nu} \sim r^{- \alpha_{\mu \nu}}$, where $\mu ,\nu$ takes values in $\{0, x,y,z\}$. For the correlations that vanish, we assume that the corresponding decay exponent $\alpha$ is infinite. Thus, we can write the correlation function as:
\begin{equation}
\text{Tr}(\rho_{1, r} \sigma_{\mu}^{1} \sigma_{\nu}^{r}) \sim r^{-\alpha_{\mu \nu}} + c_{\mu \nu},
\end{equation}
where $c_{\mu \nu} \equiv \text{Tr}(\rho_1 \sigma_{\mu}^{1}) \text{Tr}(\rho_r \sigma_{\nu}^{r})$ is independent of $r$ due to translational invariance. Using this, the total nonstabilizerness of the state can be expressed as:
\begin{equation}
\begin{aligned}
&M(\rho_{1,r}) \sim \text{ln}\left(\frac{\sum_{\mu \nu}(r^{-\alpha_{\mu \nu}}+c_{\mu \nu})^2}{\sum_{\mu \nu}(r^{-\alpha_{\mu \nu}}+c_{\mu \nu})^4}\right) \\
&\sim \text{ln}\left(\frac{c_1 + \sum_{\mu \nu}2c_{\mu \nu}r^{-\alpha_{\mu\nu}}+\sum_{\mu \nu}r^{-2\alpha_{\mu \nu}}}{c_2 + \sum_{\mu \nu}4c_{\mu \nu}^3r^{-\alpha_{\mu \nu}}+\sum_{\mu \nu}6c_{\mu \nu}^2r^{-2\alpha_{\mu \nu}}+\dots }\right),
\end{aligned}
\end{equation}
where $c_1$ and $c_2$ represent the summations of the constant terms that are independent of $r$ in the numerator and denominator, respectively. The ellipses indicate terms that decay with higher exponents. In the large $r$ limit, we can further approximate as:
\begin{equation}
\begin{aligned}
&M(\rho_{1,r}) \sim \text{ln}\left(\frac{c_1}{c_2}\right)\\
&+\left[1+\frac{1}{c_1}(\sum_{\mu \nu}2c_{\mu \nu}r^{-\alpha_{\mu\nu}}+\sum_{\mu \nu}r^{-2\alpha_{\mu \nu}})\right] \times\\
&\left[1-\frac{1}{c_2}(\sum_{\mu \nu}4c_{\mu \nu}^3r^{-\alpha_{\mu \nu}}+\sum_{\mu \nu}6c_{\mu \nu}^2r^{-2\alpha_{\mu \nu}}+\dots )\right],
\end{aligned}
\end{equation}
where $(1+x)^{-1}\sim (1-x)$ and $\text{ln}(1+x)\sim x$ for small $x$ are used. Denoting $\alpha^{*} \equiv  \min_{\mu \nu} \alpha_{\mu \nu} $ and $\overline{\alpha} \equiv  \min_{\mu \nu} \{ \alpha_{\mu \nu}; c_{\mu \nu} \neq 0\} $, it follows that the in the large $r$ limit, the total nonstabilizerness $M$ is governed by:
\begin{equation}
M(\rho_{1,r}) \sim \text{const} + r^{-2\alpha^{*}} + r^{-\overline{\alpha}}.
\end{equation}
Thus, the decay exponent of the total nonstabilizerness is given by:
\begin{equation}
\alpha_{M} = \text{min}(2\alpha^{*}, \overline{\alpha}).
\end{equation}
This conclusion holds generally for any particular critical state.  

At the critical point of the TFIM, we have $\alpha^{*} = \alpha_{xx} = 1/4$ and $\overline{\alpha} = \alpha_{zz} = 2$~\cite{francesco2012}, which results in $\alpha_{M} = 1/2$, identical to the decay exponent of the mutual information. Furthermore, since the ground state is in canonical form, the NN is equivalent to the total nonstabilizerness of the two-qubit state. This completes the proof.

\section{Numerical details}
\label{numerical}
In this section, we provide numerical details for the simulations conducted in the main text. The ground state of the TFIM is obtained using the density matrix renormalization group algorithm using the ITensor package~\cite{itensor, itensor-r0.3}, with the maximum bond dimension $\chi = 500$ and the truncation cutoff $\epsilon = 10^{-12}$. For the monitored Haar-random unitary circuit at the critical point, the time evolution was performed by the time-evolving block decimation algorithm, with the maximum bond dimension $\chi = 1000$ and the truncation cutoff $\epsilon=10^{-6}$. 

When calculating the NN via numerical optimization over all local unitary transformations, we parameterize the local unitary on a single qubit as $ U = \begin{pmatrix}
 \text{cos}(\theta/2) & -e^{i\lambda}\text{sin}(\theta/2) \\
 e^{i\phi} \text{sin}(\theta/2) & e^{i(\phi+\lambda)}\text{cos}(\theta/2)
\end{pmatrix}$, where $\phi \in [0, 2\pi]$, $\theta \in [0, \pi]$ and $\lambda \in [0, 2\pi]$. The optimization process employed the Nelder-Mead algorithm, as implemented in the Optim package~\cite{mogensen2018}. To mitigate the risk of converging to a local minimum, we initiated the optimization from 100 distinct starting points and selected the minimal value achieved.

\section{Nonzero NN in separable mixed states}
\label{separable_mixed}
In this section, we demonstrate that for any choice of nonstabilizerness measure for mixed states, certain separable states will always exhibit a nonzero NN. A separable state is defined as the convex hull of pure product states~\cite{horodecki2009}:
\begin{equation}
\label{supp:eq4}
\rho = \sum_i p_i \rho_{iA} \otimes \rho_{iB}.
\end{equation}
Within the framework of resource theory, there are two inequivalent ways to define mixed stabilizer states, which we denote as $\text{STAB}_0$ and $\text{STAB}$, following the notation in Ref.~\cite{cao2024a}. 
$\text{STAB}_0$ is defined as the set of states $\rho = \frac{1}{2^{N}}\sum_{P \in G} P$, where $G$ is a group of commuting Pauli operators $P$~\cite{nielsen2010}, while $\text{STAB}$ is defined as the convex hull of pure stabilizer states:
\begin{equation}
\label{supp:eq5}
\text{STAB} := \{\rho | \rho = \sum_i p_i \left | \psi_i \right \rangle \left \langle \psi_i \right |\},
\end{equation}
where $\psi_i$ are pure stabilizer states. When $\text{STAB}$ is regarded as the set of free states, it specifically captures distillable non-Clifford resources. On the other hand, regarding $\text{STAB}_0$ as free states, it also incorporates the non-flat probability distributions of mixed states as resources. In the second SRE used in the main text, $\text{STAB}_0$ is treated as the set of free states, while other nonstabilizerness measures, such as the robustness of magic (RoM)~\cite{howard2017, ahmadi2018,heinrich2019} and mana~\cite{veitch2014}, consider $\text{STAB}$ as the set of free states. It is also worth mentioning that the SRE has been extended by a convex-roof construction to provide an alternative measure in the latter case~\cite{leone2024b}.

As outlined in the main text, unitary transformations cannot transform even a single-qubit state into one with a flat probability distribution, i.e., $ \min_{U}M(U\rho U^{\dagger})$ is not necessarily zero. This directly implies that for a separable state $\rho_A \otimes \rho_B$, we have:
\begin{equation}
 \mathcal{M}(\rho_A \otimes \rho_B)  = \min_{U_A}M(\rho_A) + \min_{U_B}M(\rho_B) \neq 0.
\end{equation}
This scenario changes when $\text{STAB}$ is considered as the set of free states, since any single-qubit density matrix can always be diagonalized, thus ensuring that it lies within $\text{STAB}$. However, we are unaware of any proof guaranteeing that every separable multi-qubit state can be transformed into a state belonging to $\text{STAB}$ using only local unitary transformations, and we conjecture that this is not the case. To provide numerical evidence, consider the separable state:
\begin{equation}
 \rho_0 = \frac{1}{2}\left | \phi_0 \phi_0 \right \rangle \left \langle \phi_0 \phi_0 \right | + \frac{1}{2}\left | 00 \right \rangle \left \langle 00 \right |.
\end{equation}
Here, $ \left | \phi_0 \right \rangle = \text{cos}(\pi/8)\left | 0 \right \rangle + \text{sin}(\pi/8)\left | 1 \right \rangle$. Using RoM as the nonstabilizerness measure, which is defined as $R(\rho) = \min_x\{\sum_i |x_i|; \rho = \sum_i x_i \left | \psi_i \right \rangle \left \langle \psi_i \right |\}-1$, we numerically minimize the nonstabilizerness over all possible local unitary transformations and find that $\mathcal{M}_R(\rho_0) \approx 0.0703$, which is nonzero. This indicates that even when using nonstabilizerness measures that treat \text{STAB} as the set of free states, certain separable states can still exhibit a nonzero NN.

\end{appendix}

% \bibliography{refs}

\begin{thebibliography}{85}
\bibitem [{mix()}]{mixed}%
  \BibitemOpen
  \href@noop {} {\enquote {\bibinfo {title} {We emphasize that if the convex hull of pure stabilizer states is considered resource-free, then the second {{SRE}} is no longer faithful.}}\ }\BibitemShut {NoStop}%
\end{thebibliography}

\end{document}